\documentstyle{article}
\begin{document}
\renewcommand{\baselinestretch} {1.5}
\large
\hyphenation{anti-ferro-mag-netic}

\vskip 1.0in
\begin{center}
{\large{\bf An Improvement on the Negative Sign Problem\\
in Numerical Calculations of a Quantum Spin System\\}}
\vskip 1.0in
   Tomo Munehisa and Yasuko Munehisa\\
\vskip 0.5in
   Faculty of Engineering, Yamanashi University\\
   Kofu, Yamanashi, 400 Japan\\
\vskip 1.5in
\end{center}

We propose new approach to numerical study of quantum spin systems.
Our method is based on a fact that one can use any set of states
for the path integral as long as it is complete.
We apply our method to one-dimensional quantum spin system
with next-to-nearest neighbor interactions.
We found remarkable improvement in negative sign problem.

\eject
\noindent {\bf Section 1  Introduction}

Recently quantum spin systems have obtained much interests
among people in various fields.
One reason for it is that Haldane found characteristic
property of the quantum spin system which is difficult to
imagine in the classical case \cite{fdmh}.
Another reason comes from the possible relationship between
the antiferromagnetic system on a 2-dimensional square lattice
and the high $T_c$ materials \cite{hightc}.

One powerful tool to numerically investigate quantum spin systems
is Monte Carlo approach using the Suzuki-Trotter formula \cite{st}.
Study through this method has brought us very intriguing
results on the ferromagnetic system.
If one applies this method to the antiferromagnetic system,
however, one often encounters the so-called negative sign
(N.S.) problem.
N.S. problem, which becomes more serious on larger lattices,
makes it very difficult to get
statistically meaningful results in numerical calculations.

Let us explain what the N.S. problem is.
To this purpose we show how the Monte Carlo method is
applied to the quantum spin $1/2$ system whose Hamiltonian is
  $$   \hat{H} = \sum \vec{\sigma}_i \vec{\sigma}_j, $$
where $\vec{\sigma}_i=(\sigma_i^x,\sigma_i^y,\sigma_i^z)$
represents Pauli matrix on site $i$ of a
lattice and sum runs over all nearest neighbors.
The partition function $Z$ is calculated by
  $$ Z = tr( e^{-\beta \hat{H} } ) $$
Using the Suzuki-Trotter formula it can be written by the
classical partition function as follows.
  $$ Z = \lim_{n \rightarrow \infty}
     tr \lbrace (e^{-\beta \hat{H}_1/n }
     e^{-\beta \hat{H}_2/n } ) ^n \rbrace $$
  $$ = \lim_{n \rightarrow \infty}
     \sum_{ \lbrace \alpha_i, \alpha'_i \rbrace } \prod_{i=1}^n
    < \alpha_i \mid e^{-\beta \hat{H}_1/n} \mid \alpha'_i >
    < \alpha'_i \mid  e^{-\beta \hat{H}_2/n} \mid \alpha_{i+1} >, $$
  $$  \hat{H} = \hat{H}_1 + \hat{H}_2. $$
Here $\mid \alpha_i>$ represents a state of the system,
$\lbrace \mid \alpha_i> \rbrace $ a complete set of states
  $$  \hat{1} = \sum_{\alpha_i} \mid \alpha_i > < \alpha_i \mid, $$
and $\hat{H}$ is divided into two parts so that
every term in $\hat{H}_1$ ($\hat{H}_2$) commutes
with other terms within that partial Hamiltonian.
Since the expectation value of $e^{-\beta \hat{H}/n}$ is
a real number we can apply Monte Carlo methods to
this partition function if $n$, which is called trotter number,
is kept finite.

There would have been no N.S. problem
if positivity of the expectation value were always guaranteed.
However in several cases of much physical interest,
antiferromagnetic quantum spin system for example,
we do have some negative expectation values. If total product
of expectation values over one configuration can become negative,
then the N.S. problem may occur. In order to numerically calculate
some physical quantity $\langle A \rangle $ one should subtract
contributions of negatively signed configurations, $A_{-}$,
from those of positively signed ones, $A_{+}$, namely,
  $$ \langle A \rangle ={{A_{+}-A_{-}} \over {Z_{+}-Z_{-}}},$$
where $Z_{+}(Z_{-})$ is number of configurations with positive
(negative) weight. The result would suffer from serious
cancellation when $Z_{-} \simeq Z_{+}$.

Note that negativity of expectation values does not always
bring the N.S. problem. On square lattices one does
not need to worry about it even in the antiferromagnetic case
because the total product of expectation values is always
positive. On the triangular lattice, on the contrary,
the N.S. problem is quite serious.

In this paper we describe our prescription to obtain
meaningful numerical results for systems to which the
conventional formulation is useless because of the N.S. problem.
Our strategy is to find rearranged basis of the system
with which no serious cancellations take place.
We do not mean that the N.S. problem is completely solved by our
approach, but we show our method is successful to improve the
numerical results when applied to a quantum spin system.

In following sections we concentrate ourselves to one dimensional
quantum spin $1/2$ system with the next-to-nearest neighbor
interactions, which is the simplest one among systems suffering
from serious N.S. problem.
We show what choice of complete set of states is more suitable
in this case to carry out Monte Carlo simulations.
In section 2 we present our method.
Numerical results will be given in section 3 and final section is
devoted to discussions on technical problems.
Physical properties of the system will be discussed in other paper.

\vskip 0.5in
\noindent {\bf Section 2   Method}

Hamiltonian of the quantum spin $1/2$ system with next-to-nearest
neighbor interactions on a one-dimensional chain
is
  $$  \hat{H}= -{J \over 2} \sum_{i=1}^N
     (\vec{\sigma}_i\vec{\sigma}_{i+1} +
      \vec{\sigma}_i\vec{\sigma}_{i+2}), $$
where $N$ is number of sites on the chain
and $\vec{\sigma}_{N+i} \equiv \vec{\sigma}_i$
(periodic boundary condition).
The system is ferromagnetic (antiferromagnetic)
for positive (negative) $J$.
We imply the coupling constant between next-to-nearest neighbors
is equal to that between nearest neighbors. It is partly because
for simplicity and partly because this case brings most serious
N.S. problem for negative $J$.
The partition function $Z$ is given by $Z=tr(e^{-\beta \hat{H}})$.

First let us describe the conventional approach. State on each site
is represented by $z$-component of the spin, namely up and down,
or $+$ and $-$.
In this representation states of the system are given by
  $$    \mid \alpha > = \mid s_1,s_2,s_3,...,s_N > , $$
where $s_i=+$ or $-$. The identity operator is then
  $$  \hat{1} = \sum_{\lbrace s_i \rbrace} \mid s_1,s_2,...,s_N >
      < s_1,s_2,...,s_N \mid . $$
To use this identity operator in the Suzuki-Trotter formula,
we divide the Hamiltonian into four parts as schematically
shown in Fig. 1(a).
  $$ Z = \lim_{n \rightarrow \infty}
     tr \lbrace (e^{-\beta \hat{H}_1/n }
     e^{-\beta \hat{H}_2/n }
     e^{-\beta \hat{H}_3/n }
     e^{-\beta \hat{H}_4/n } ) ^n \rbrace ,$$
   $$  \hat{H}_1 = -{J \over 2} \sum_{i=1}^{N/2}
       \vec{\sigma}_{2i-1}\vec{\sigma}_{2i}, $$
   $$  \hat{H}_2 = -{J \over 2} \sum_{i=1}^{N/2}
       \vec{\sigma}_{2i}\vec{\sigma}_{2i+1}, $$
   $$  \hat{H}_3 = -{J \over 2} \sum_{i=1}^{N/4}
       (\vec{\sigma}_{4i-3}\vec{\sigma}_{4i-1} +
        \vec{\sigma}_{4i-2}\vec{\sigma}_{4i}), $$
   $$  \hat{H}_4 = -{J \over 2} \sum_{i=1}^{N/4}
       (\vec{\sigma}_{4i-1}\vec{\sigma}_{4i+1} +
        \vec{\sigma}_{4i  }\vec{\sigma}_{4i+2}). $$
With these complete sets and partial Hamiltonians we obtain the
partition function $Z_C^{(n)}$ which we use in Monte Carlo
calculations,
   $$ Z_C^{(n)} = \sum_{\lbrace \alpha_j, \alpha'_j, \alpha''_j,
      \alpha'''_j \rbrace } \prod_{j=1}^n
      <\alpha_j \mid e^{-\beta\hat{H}_1/n} \mid \alpha'_j>
      <\alpha'_j \mid e^{-\beta\hat{H}_2/n} \mid \alpha''_j> $$
   $$ \ \ \ \ \ \ \times
      <\alpha''_j \mid e^{-\beta\hat{H}_3/n} \mid \alpha'''_j>
      <\alpha'''_j \mid e^{-\beta \hat{H}_4/n} \mid \alpha_{j+1}>
      , $$
where suffix $j$ numbers sites along the trotter axis
and $\alpha_{n+1} \equiv \alpha_1$.

Now we turn to our method. First we will show our idea using spin
system of two sites which we denote site $a$ and site $b$.
The Hamiltonian of this system is
simply $\hat{H}_{ab}= -J \vec{\sigma_a} \vec{\sigma_b} $.
In the conventional approach to this two-spin system
the complete set is given by
$ \lbrace \mid +_a,+_b >,  \mid +_a,-_b >, \mid -_a,+_b >,
\mid -_a,-_b > \rbrace .$
Instead of this set we can adopt another complete set composed of
diagonalized states of $\hat H_{ab}$,
   $$ \lbrace \mid 1 >,  \mid \oplus >, \mid \ominus >, \mid -1 >
      \rbrace ,$$
where
   $$ \mid 1 > =\mid +_a , +_b > ,$$
   $$ \mid \oplus > = {1\over \sqrt{2} }
     (\mid +_a , -_b >+ (\mid -_a , +_b >),$$
   $$ \mid \ominus > ={1\over\sqrt{2} }
     (\mid +_a , -_b >- (\mid -_a , +_b >),$$
   $$ \mid -1 > = \mid -_a , -_b >  . $$
The partition function is then easily calculated using eigenvalues
$E_1$, $E_\oplus$, $E_\ominus$ and $E_{-1}$,
   $$ Z_{ab}=tr( e^{-\beta \hat{H}_{ab}})$$
   $$ = e^{-\beta E_1}+e^{-\beta E_{\oplus}}+ e^{-\beta E_{\ominus}}+
      e^{-\beta E_{-1}}.$$
We apply this diagonalization to the spin system of $N$ sites.
For this purpose we rewrite the Hamiltonian in the following form
   $$  \hat{H}= -{J \over 2} \sum_{i=1}^{N/2}
      (\vec{\sigma}_{a,i}\vec{\sigma}_{a,i+1}  +
       \vec{\sigma}_{b,i}\vec{\sigma}_{b,i+1}  +
       \vec{\sigma}_{a,i}\vec{\sigma}_{b,i}    +
       \vec{\sigma}_{b,i}\vec{\sigma}_{a,i+1}), $$
where we denote odd and even sites with suffix $a$ and $b$,
respectively,
   $$\vec{\sigma}_{a,i} \equiv \vec{\sigma}_{2i-1},\ \
     \vec{\sigma}_{b,i} \equiv \vec{\sigma}_{2i}.$$
We employ the complete set where the operators
$\vec{\sigma}_{a,i} \vec{\sigma}_{b,i} $ are diagonalized.
Then each state of the system is represented by combinations
of $N/2$ two-site diagonalized states,
  $$ \mid \alpha > = \mid S_1,S_2,...,S_{N/2} >, $$
where $S_i$ stands for diagonalized state on the $i$-th spin
pair, namely $S_i=1_i$, $\oplus_i$, $\ominus_i$ or $-1_i$.
Then it becomes necessary to divide the Hamiltonian
into ``odd'' and ``even'' parts
  $$ \hat{H}=\hat{H_o}+\hat{H_e}, $$
where, as shown in Fig. 1(b),
  $$ \hat{H_o} = -{J \over 2} \sum_{i=1}^{N/4}
     (\vec{\sigma}_{a,2i-1} \vec\sigma_{a,2i}
     + \vec\sigma_{b,2i-1} \vec\sigma_{b,2i}
     + \vec\sigma_{b,2i-1} \vec\sigma_{a,2i} $$
  $$ \ \ \ \ \ + {1 \over 2} \vec\sigma_{a,2i-1} \vec\sigma_{b,2i-1}
     + {1 \over 2} \vec\sigma_{a,2i} \vec\sigma_{b,2i} ), $$
  $$ \hat{H_e} = -{J \over 2} \sum_{i=1}^{N/4}
     (\vec{\sigma}_{a,2i} \vec\sigma_{a,2i+1}
     + \vec\sigma_{b,2i} \vec\sigma_{b,2i+1}
     + \vec\sigma_{b,2i} \vec\sigma_{a,2i+1} $$
  $$ \ \ \ \ \ + {1 \over 2} \vec\sigma_{a,2i-1} \vec\sigma_{b,2i-1}
     + {1 \over 2} \vec\sigma_{a,2i} \vec\sigma_{b,2i} ). $$
The partition function with these partial hamiltonians is
  $$   Z=\lim_{n \rightarrow \infty} tr \lbrace
      (e^{-\beta \hat{H}_o/n} e^{-\beta \hat{H}_e/n} )^n \rbrace . $$
Inserting identity operators made of $\alpha$'s between the
exponents we obtain partition function $Z_M^{(n)}$ to use in
our numerical study,
  $$ Z_M^{(n)} = \sum_{ \lbrace \alpha_j,\alpha'_j \rbrace }
     \prod_{j=1}^{n} <\alpha_j \mid e^{-\beta \hat{H}_o/n}
     \mid \alpha'_{j} > <\alpha'_j \mid
     e^{-\beta \hat{H}_e/n} \mid \alpha_{j+1}>. $$

\eject
\noindent {\bf Section 3   Numerical Results}

In this section we present Monte Carlo results we obtained
using the new partition function stated in section 2.
In order to figure how much quantitative improvement we
can make, we also show some results from the conventional
approach.

Monte Carlo simulations are carried out on a two dimensional
lattice, one of which is space direction of size $N$ and
another is the trotter direction having $2n$ ($4n$) sites
in our (conventional) method with the trotter number $n$.

In each simulation with new partition function $Z_{M}^{(n)}$
typically one thousand configurations are generated
in order to reach thermal equilibrium and ten thousand
configurations after thermalization are used
to measure system's energy, $E$, and ratio of negatively
weighted configurations to total configurations, $P$,
   $$ E=-{{\partial} \over {\partial \beta}} ln Z, $$
   $$ P={{Z_-} \over {Z_+ + Z_-}}. $$
Here one new configuration is obtained by updating spins on
all sites using the heat-bath method. In each update
we make simultaneous spin flips on sites along the smallest
closed path (local flips) and along the trotter axis
($n$-direction global flips) according to their probabilities.
{}From technical reasons we did not include local and global flips
which change $z$-component of $S_i$ of $i$-th spin pair by 2 \cite{fn1}.
We also made global flips along the space axis ($x$-direction
global flips) available, but we excluded them in simulations presented
in this paper because we have found their effects on system's
energy are negligible.

In simulations with the conventional partition function $Z_{C}^{(n)}$
we employ Metropolis algorithm. Here,
in addition to the local flips and usual $n$-direction global flips,
we found it necessary to make a kind of twisted $n$-direction global
flips shown in Fig. 1(c) so that enough phase space is guaranteed.
We carry out twenty thousand updates for the thermalization, one hundred
thousand for the measurement.

Let us first present results for the ferromagnetic system.
In this case the conventional method should be effective
since no N.S. problem exists.
Fig. 2 shows results on an $N=8$ chain by the conventional
method as a function of temperature $T$,
which is defined by $1/J \beta$, together with exact values
obtained by diagonalization methods.
We see conventional method works very well.
Our method is also successful, but we have no special reason
to prefer it \cite {fn2}.

Next we will show results for the antiferromagnetic system.
In Fig. 3(a) we present ratio $P$ versus $T$ with $N$=8 chain by
both methods.
As we see in the figure, even when $n=2$ the ratio $P$ with the
conventional partition function increases quite rapidly
as $T$ decreases. It is therefore difficult to obtain
statistically meaningful results by this method.
By our new method, on the other hand, $P$ remains much smaller
for larger $n$ and for lower $T$ so that the N.S. problem is
less serious.
Fig. 3(b) plots system's energy on $N=8$ and $n=2$, 3 and 4
lattices in comparison with exact values.
The results indicate we can get reliable values by new method.

How far is our method useful when the size of the chain
is enlarged? Fig. 3(c), which shows how ratio $P$ in our method
increases for larger $N$'s, answers this question.
We see that with our present choice of states reliable calculation
on $N \geq 32$ and $n \geq 4$ lattices is difficult when $T < 1.0$.

\vskip 0.5in
\noindent {\bf Section 4   Discussions}

In this paper we suggested a new approach for Monte Carlo study
of quantum spin systems suffering from the negative sign problem.
Essential point of our method is to choose a set of states for the
path integral which is appropriate to numerical calculation.
Conventional choice of the set, whose states consist of eigenstates of
$z$-component of the Pauli matrix on each lattice site, is the simplest
one but the N.S. problem turns out to be serious with this set.
Since any other choice is possible as long as the set is complete,
we employ a set made of eigenstates of diagonalized Hamiltonian for
every two neighboring sites. We believe this choice is better
approximation to true ground state and excited states.

We applied our method to one-dimensional quantum spin $1/2$ system
with next-to-nearest neighbor interactions, which suffers from the
N.S. problem in the antiferromagnetic case.
Then remarkable improvements were found as shown in Section 3.
Monte Carlo results also showed, however, that even in our method
one would encounter difficulties for low $T$ on large lattices
because the ratio of negatively signed configurations increases
up to $\sim$ 0.5 when temperature goes down or lattice size is
enlarged.
So we should emphasize that what we present here is not a way to
completely solve the N.S. problem but a prescription to improve
numerical calculations.

Comments on ergodicity are in order. In quantum Monte Carlo
simulations it is known that local flips are not sufficient to guarantee
ergodic update of the system \cite{erg}. One should add some kinds of
global spin flips --- $n$-direction global flips to change $z$-component
of spins, $x$-direction global flips and transverse global flips to
change the winding number, and so on --- in updating procedure in
order to make phase space big enough to observe satisfactorily precise
results. What global flips are necessary and sufficient is not a trivial
question, but from technical point of view it is preferred to include
minimum kinds of global flips which are as simple as possible.
{}From our experiences $n$-direction global flips seem inevitable
for this purpose.
As results in Section 3 indicate, we can measure system's energy
without any other flips in wide range of temperature.
Transverse global flips and $x$-direction global flips, however,
might be important for low temperatures. It should be investigated
in future study.

Although a complete solution for N.S. problem could not be obtained,
our results show much improvement in numerical study, which inspires
us with confidence that quantum Monte Carlo methods are effective
if a set of states for the path integral is appropriately chosen.
More improvements would be possible by selecting larger cluster of
spins to construct states which diagonalize the Hamiltonian.

\eject

\eject
\noindent {\bf Figure captions}

\bigskip
\noindent {\bf Fig. 1}

(a) Way to divide Hamiltonian on $N$=8 chain in the conventional
approach.
Filled circles, squares, diamonds and triangles indicate terms
in $\hat{H}_1, \hat{H}_2, \hat{H}_3$ and $\hat{H}_4$, respectively.

(b) Way to divide Hamiltonian on $N$=8 chain in our approach,
where odd and even sites are renumbered as shown.
Open triangles and diamonds represent terms in $\hat{H}_1$ and
$\hat{H}_2$, respectively.
Open circles denote terms halved by $\hat{H}_1$ and $\hat{H}_2$.

(c) Examples for twisted $n$-direction global flips when $N=8$.
Open circles and plaquettes with vertical trotter direction
denote site locations and elementary
four spin interactions, respectively.
All spins on each path shown by bold solid line should be
simultaneously flipped.

\bigskip
\noindent {\bf Fig. 2}

Energy in the ferromagnetic case versus temperature $T$.
Data are calculated on $N$=8 chain by diagonalization method (exact)
and by the conventional approach for trotter number $n=2$, 3 and 4.
For each $T$ and $n$ twenty thousand sweeps are done for the
thermalization, one hundred thousand for measurement.

\bigskip
\noindent {\bf Fig. 3}

Data obtained by new method in the antiferromagnetic case.
Each measurement is done using ten thousand configurations generated
after one thousand sweeps for the thermalization.

(a) Ratio of negatively signed configurations on $N$=8 chain
as a function of temperature.
Results from conventional approach are also plotted for comparison.
In simulations by conventional method twenty thousand sweeps are done
for the thermalization, one hundred thousand for measurement.

(b) Energy of the system observed on $N$=8 and $n$=2, 3 and 4 lattices.

(c) Ratio of negatively signed configurations
as a function of the number of lattice sites.
Trotter number $n$ is kept 4 here.


\begin{thebibliography}{99}
\bibitem{fn1}
Contribution from $n$-direction global flips with $\Delta S_i^z = 2$
should be estimated in future work because it is not clear whether two
successive $\Delta S_i^z = 1$ updates can realize all $\Delta S_i^z=2 $
configurations.
\bibitem{fn2}
It should be noted that in our method
configurations with negative weights do appear even in the
ferromagnetic system, though they are small in number.
\end{thebibliography}
\end{document}